\documentclass[twocolumn,showpacs,preprintnumbers,prb,amsmath,amssymb]{revtex4}

\usepackage[dvips]{graphicx}
\usepackage{dcolumn}
\usepackage{bm}
\usepackage{amsmath}

\begin{document}

\title{Fano Resonance in a Quantum Wire with a Side-coupled Quantum Dot}

\author{Kensuke Kobayashi, Hisashi Aikawa, Akira Sano} 
\author{Shingo Katsumoto}
\author{Yasuhiro Iye}

\affiliation{ Institute for Solid State Physics, University of Tokyo,
5-1-5 Kashiwanoha, Chiba 277-8581, Japan}

\date{\today}

\begin{abstract}
We report a transport experiment on the Fano effect in a quantum
connecting wire (QW) with a side-coupled quantum dot (QD).  The Fano
resonance occurs between the QD and the ``T-shaped'' junction in the
wire, and the transport detects antiresonance or a forward scattered
part of the wave function. While it is more difficult to tune the
shape of the resonance in this geometry than in the previously
reported Aharonov-Bohm-ring-type interferometer, the resonance purely
consists of the coherent part of transport.  Utilizing this advantage,
we have quantitatively analyzed the temperature dependence of the Fano
effect by including the thermal broadening and the decoherence. We
have also proven that this geometry can be a useful interferometer for
measuring the phase evolution of electrons at a QD.
\end{abstract}
\pacs{PACS numbers: 73.21.La, 85.35.-p, 73.23.Hk, 72.15.Qm}


\maketitle

\section{Introduction}
The first observation of coherent transport in a mesoscopic
system\cite{Webb1985} opened up the field of electron interferometry
in solids.  Following the development of the Aharonov-Bohm (AB)-type
interferometer, there appeared various types of electron
interferometers including the Fabry-P{\'e}rot type~\cite{Smith1989}
and the Mach-Zehnder type~\cite{JiNature2003}.  Such interferometry is
of particular interest when the propagating electron experiences
electronic states in a quantum dot (QD), because the interference
pattern provides information on the physical properties of the QD, for
example, the electron correlation inside it.  Several interferometry
experiments have been reported for a QD embedded in an AB
ring~\cite{YacobyPRL1995,kats1996,SchusterNature1997,%
vanderWielScience2000,JiScience2000,JiPRL2002}.

\begin{figure}[tb]
\center \includegraphics[width=0.75\linewidth]{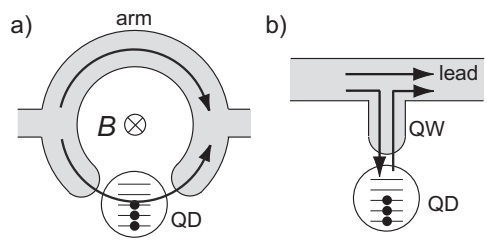}
\caption{(a) Schematic of a Fano system consisting of a quantum dot (QD)
and an AB ring.  (b) Another type of Fano system consisting of a
connecting quantum wire (QW) with a ``T-coupled'' QD. }
\label{FanoSchematic}
\end{figure}

In these experiments one should remember that the unitarity of
electron wave propagation inevitably affects the transport property of
the system. In the case of two-terminal devices, for example, this
results in the phase jump of AB oscillation at the
resonance~\cite{YeyatiPRB1995}.  While the resonance has such a subtle
aspect, it brings interesting effects on the transport when it is
positively used.  A representative is the Fano
effect~\cite{FanoPR1961}, which appears as a result of interference
between the localized state and the continuum.  Although the Fano
effect has been established in spectroscopy~\cite{FanoRau1986etc}, its
general importance in mesoscopic transport has been recognized only
recently~\cite{TekmanPRB1993,NockelPRB1994}.  It has been predicted
that the Fano effect appears in a QD embedded in an AB ring as
schematically shown in
Fig.~\ref{FanoSchematic}~(a)~\cite{DeoMPL1996etc}, and recently, we
have reported on its first experimental
observation~\cite{KobayashiPRL2002,KobayashiPRB2003}.

The AB geometry has an advantage such that the interference pattern
can be tuned by the magnetic flux piercing the ring, while its spatial
size tends to be large.  In order to maintain the quantum coherence
and to observe clearer effects, a smaller scale interferometer would
be desirable.  A candidate is the quantum wire (QW) with a finite
length.  A QD plus a connecting QW with measurement leads can be
realized in the system schematically shown in
Fig.~\ref{FanoSchematic}~(b).  Here, the very short QW connecting the
QD and the lead works as a resonator.  We call this geometry a
``side-coupled'' QD or a ``T-coupled'' QD.

The Fano effect is expected to occur in the T-coupled QD in a way
different from that in the QD-AB-ring system, because only reflected
electrons at the QD are involved for its emergence. In the Fano effect
in the QD-AB-ring system reported
previously~\cite{KobayashiPRL2002,KobayashiPRB2003}, the transmission
through the QD played a central role as manifested in the amplitude of
the AB oscillation that was in the same order of the Coulomb peak
height. Thus, two types of Fano effect can be defined as ``reflection
mode'' and as ``transmission mode''.  In general, both modes coherently
exist in the QD-AB-ring system (a typical example will be discussed
below in Sec.~\ref{subsec2}.). On the other hand, only the reflection
mode features in the T-coupled QD.  While the Fano effect in the
reflection mode has been discussed
theoretically~\cite{KangPRB2001,ClerkPRL2001,
DavidovichPRB2002,AligiaPRB2002,TorioPRB2002,FrancoPRB2003,KonikCM2002},
the experimental realization has been lacking.  Furthermore, although
the reflection amplitude itself conveys rich information on the QD, it
has not fully been investigated since the pioneering experiment by Buks
\textit{et al}~\cite{BuksPRL1996}.

In this paper, we report on the first experimental observation of the
Fano effect in a T-coupled QD.  After describing the experimental
setup in Sec.~\ref{secExp}, evidence for the emergence of the Fano
state with decreasing temperature is given in Sec.~\ref{subsec1}.  We
discuss the temperature dependence of the coherence measured in this
geometry in Sec.~\ref{SecTdep}.  Then in Sec.~\ref{subsec2}, we show
that the Fano effect in a T-coupled QD can be used to detect the phase
shift in the scattering by the QD, which makes it a unique tool for
investigating the phase and coherence of electrons in a QD.

\section{Experiment}
\label{secExp}
\begin{figure}[bth]
\center \includegraphics[width=0.6\linewidth]{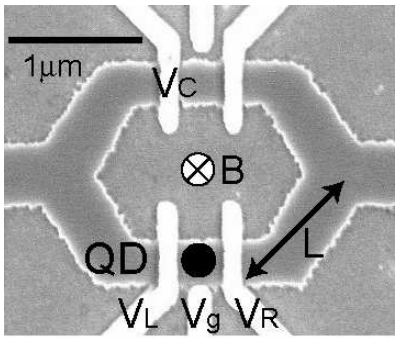}
\caption{Scanning electron photomicrograph of the device fabricated by
wetetching the 2DEG at an AlGaAs/GaAs heterostructure.  The three
gates (to which voltages $V_L$, $V_g$, and $V_R$ are applied as
indicated in the figure) on the lower arm are used for controlling the
QD, and the gate voltage $V_C$ is used for tuning the conductance of
the upper arm. When the gate $V_L$ is biased strongly, the system
becomes a T-coupled QD, as shown in Fig.~\ref{FanoSchematic} (b). The
length of the QW, namely, the distance between the QD and the junction
at the lead, is approximately $L \sim 1~\mu$m.}
\label{SampleFig}
\end{figure}

To realize a T-coupled QD, we fabricated the device shown in the
scanning electron photomicrograph of Fig.~\ref{SampleFig}.  It was
fabricated from an AlGaAs/GaAs heterostructure by wetetching. The
characteristics of the two-dimensional electron gas (2DEG) were as
follows: mobility $=9\times 10^5$~cm$^2/$Vs, sheet carrier density
$=3.8\times10^{11}$~cm$^{-2}$, and electron mean free path $l_e =
8$~$\mu$m.  This device is similar to what we had previously
studied~\cite{KobayashiPRL2002,KobayashiPRB2003}.  Two sets of three
fingers are Au/Ti metallic gates for controlling the local
electrostatic potential.  The three gates ($V_L$, $V_g$, and $V_R$) on
the lower arm are used for defining and controlling the parameters of
the QD with a geometrical area of 0.2$\times$0.2~$\mu$m$^2$.  The gate
on the upper arm $V_C$ is used for tuning the conductance of this arm.

In the addition spectrum of the QD, the discrete energy levels inside
the QD are separated by the level spacing due to the quantum
confinement $\Delta E$ and the single-electron charging energy $E_C$.
We can shift the spectrum using the center gate voltage $V_g$ to tune
any one of them to the Fermi level.  In the present sample, by
measuring the conductance through the QD, we found that $\Delta E$ and
$E_C$ are typically $\sim 120$~$\mu$eV and $\sim 1$~meV, respectively.
We can make the device a T-coupled QD by applying a large negative
voltage on $V_L$ so that the electron transmission underneath $V_L$ is
forbidden.  This is topologically the same as the T-coupled QD shown
in Fig.~\ref{FanoSchematic}~(b).  The distance between the QD and the
junction at the lead is $L \sim 1~\mu$m.  When required, we made this
gate slightly transmissible in order to measure the AB signal through
the system with the magnetic field ($B$) applied perpendicular to the
2DEG.

Measurements were performed in a mixing chamber of a dilution
refrigerator between 30~mK and 1~K by a standard lock-in technique in
the two-terminal setup with an excitation voltage of 10~$\mu$V (80~Hz,
5~fW) between the source and the drain.  Noise filters were inserted
into every lead below $1$~K as well as at room temperature.

\section{Results and Discussion}
\subsection{Emergence of the Fano Effect in the Conductance}
\label{subsec1}

\begin{figure}[htb]
\center \includegraphics[width=0.95\linewidth]{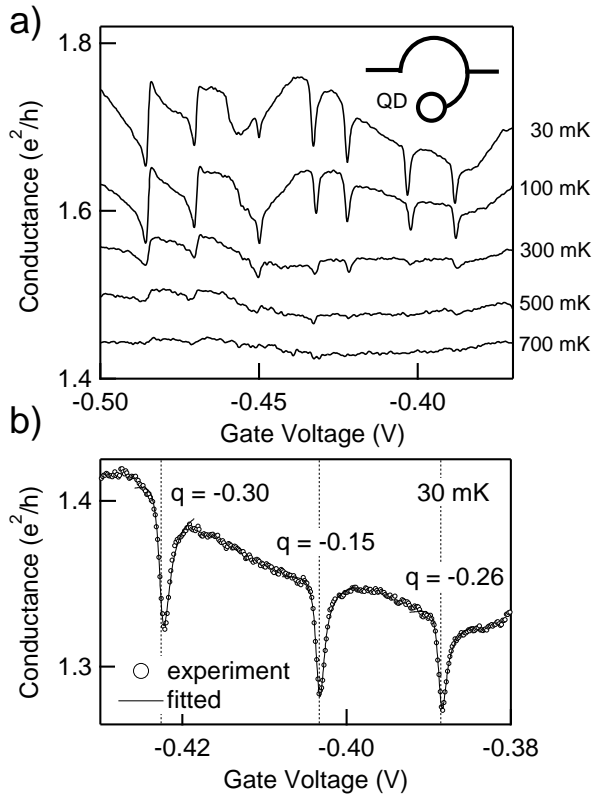}
\caption{(a) Conductance of the T-coupled QD as a function of $V_g$ at
several temperatures. As the temperature was lowered, the Fano feature
appeared. The magnetic field was 0.80~T.  The curves for $T < 700$~mK
are incrementally shifted upward for clarity. (b) Three Fano features
in the conductance at 30~mK are fitted to Eq.~\eqref{CondFanoEqn}. The
obtained $q$'s are shown. The vertical dashed lines indicate the
obtained discrete level position $V_0$'s.}
\label{TypicalData}
\end{figure}

We set $V_L$ to $-0.205$~V so as to forbid the electron transmission
under it. $V_R$ was $-0.190$~V, which made this gate slightly
transmissible.  $V_C$ was adjusted to make the conductance of the
system around $2e^2/h$, which is a quasi-single channel condition.
Figure \ref{TypicalData}~(a) shows typical results of the conductance
through the system as a function of the gate voltage $V_g$ at several
temperatures ($T$).  Since the connecting QW between the lead QW and
the QD is narrow and long, the QD would not affect the conduction
through the lead QW in the classical transport regime.
Correspondingly, at high temperatures above $800$~mK, hardly any
characteristic structure appears in the signal.  Sharp dip structures,
however, rapidly grow with decreasing temperature.  They are
antiresonance (or reflection due to resonance) dips due to Coulomb
oscillation in the QD.  Furthermore, the resonant features are very
asymmetric and vary widely in their line shape.  For example, at
$V_g=-0.485$~V and $V_g=-0.47$~V, the line shape consists of a sharp
dip and an adjacent peak, while only asymmetric sharp-dip structures
appear between $V_g = -0.45$~V and $-0.38$~V.

These line shapes in the conductance are characteristic of the Fano
effect. In fact, the line shape at the lowest temperature can be
fitted to~\cite{KobayashiPRL2002,KobayashiPRB2003}
\begin{equation} 
G_{tot} = A \frac{(\tilde\epsilon
+q)^2}{\tilde\epsilon^2+1} + G_{bg},
\label{CondFanoEqn}
\end{equation}
where $G_{bg}$ is the noninterfering contribution of the lead and is a
smooth function of $V_g$ that can be treated as a constant for each
peak.  The first term is the Fano contribution with an real asymmetric
parameter $q$ where the normalized energy
\begin{equation} 
\tilde\epsilon =\frac{\epsilon-\epsilon_{0}}{{\it
\varGamma}/2}=\frac{\alpha(V_g-V_{0})}{{\it \varGamma}/2}.
\label{CondFanoEqn2}
\end{equation}
The parameters $A$, $\epsilon_{0} \equiv \alpha V_{0}$, and ${\it
\varGamma}$ represent the amplitude, the position, and the width of
the Fano resonance, respectively. $\alpha$ is the proportionality
factor which relates $V_g$ to the electrochemical potential of the QD
and is given by $\alpha = e C_g /C_{tot}$, where $C_g$ is the
capacitance between the QD and the gate $V_g$, and $C_{tot}$ is the
total capacitance.  Note that the functional form of the Fano part in
Eq.~\eqref{CondFanoEqn} can be applied to both resonance and
antiresonance.  The parameters of the QD can be obtained from an
independent measurement of the transport through the QD under an
appropriate condition of the gate voltages (applying a large negative
$V_C$ to cut the upper conduction path and opening the lower path by
decreasing $V_L$). We estimate $\alpha =50\pm10$~$\mu$eV$/$mV for the
present system.

Figure~\ref{TypicalData}~(b) shows typical results of the fitting for
three dips. The satisfactory agreement assures that the Fano state is
formed in the T-coupled QD system.  The obtained values of ${\it
\varGamma}$ are $\sim 70$~$\mu$eV, which is much larger than the
thermal broadening $3.5k_{\rm B}T = 9$~$\mu$eV at 30~mK (here, $k_{\rm
B}$ is the Boltzmann constant).  In Fig.~\ref{TypicalData}~(b), the
obtained $q$'s are also shown and the vertical dashed lines indicate
the discrete level position ($V_0$).  Both the variation of $q$ and
that of the level spacing indicate that the conductance through this
system reflects the characteristic of each of the single levels in the
QD.

The data in Fig.~\ref{TypicalData}~(a) is obtained at $B=0.80$~T.  The
Fano effect in this system has been observed at several magnetic
fields, as was also the case in the Fano effect in the QD-AB-ring
system~\cite{KobayashiPRL2002,KobayashiPRB2003}.  In the present case,
the role of the magnetic field can be understood because the QW
between the QD and the junction is curved as shown in
Fig.~\ref{SampleFig}, and the coupling is modified by the Lorentz
force. In the previous
reports~\cite{KobayashiPRL2002,KobayashiPRB2003}, we have discussed
that $q$ should be a complex number in a QD-AB-ring geometry under
finite $B$.  This treatment is required to describe the $V_g$-$B$
dependence of the line shape to cover the wide range of $B$, when the
interfering phase is modulated by $B$.  For a fixed magnetic field,
Eq.~\eqref{CondFanoEqn} with a real $q$ well describes the line shape.
Furthermore in the present case, since the effective area of the
connecting QW is very small, the line shape is found to be much less
sensitive to the magnetic field than the case of an AB ring.

In the T-coupled geometry, the resonance is detected through the
nonlocal conductance and the electric field of the modulation gate
$V_g$ works only locally, and therefore it is easy to distinguish the
coherent part from the incoherent part in the conductance.  In
contrast, the transmission experiments such as in the QD-AB-ring
system usually provide the transmission probability including both the
coherent and incoherent processes~\cite{AikawaCM2003}.  For example,
in the case of the Fano resonance in the AB geometry, ordinary Coulomb
oscillation overlaps the coherent line shape.  In the case of simple
AB magnetoresistance, the magnetic field is applied all over the
specimen, and the AB oscillation is superposed on the background
conductance fluctuation.

\subsection{Temperature Dependence of the Fano Effect}
\label{SecTdep}
With increasing temperature, the dip structures are rapidly smeared
out.  The origins of such smearing due to finite temperature can be
classified into thermal broadening and quantum decoherence.  The
former should be considered when $3.5k_{\rm B}T$ becomes the same
order of magnitude as ${\it \varGamma}$ at $T = 200$~mK.  Henceforth,
we focus on the thermal broadening and examine whether it alone can
explain the observed diminishment of the resonance structure.

The thermal broadening appears in the distribution in $\tilde\epsilon$
in the Fano form in Eq.~\eqref{CondFanoEqn}.  As noted in the previous
paper~\cite{KobayashiPRB2003}, Eq.~\eqref{CondFanoEqn} is derived by
assuming a point interaction between the localized states and the
continuum.  However in the present system, the length of the connecting
QW is finite and dephasing during the traversal due to thermal
broadening may be important.  If such dephasing can be ignored, $q$ can
be treated as a real number as noted above.  Conversely, the dephasing
inside the connecting QW requires the use of complex
$q$'s~\cite{ClerkPRL2001}.

\begin{figure}[tb]
\center \includegraphics[width=0.95\linewidth]{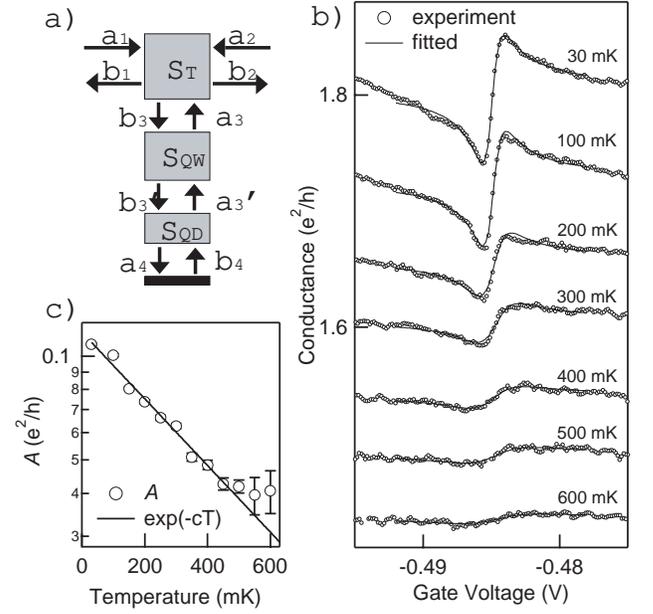}
\caption{(a) Model of the quantum circuit that consists of the
T-junction, the QW and the QD. See also Fig.~\ref{FanoSchematic} (b).
(b) Experimental data (open circles) and the fitted curves.  The data
for $T < 600$~mK are incrementally shifted upwards for clarity.  (c)
Obtained $A$ value is shown in logarithmic scale as a function of
temperature. The solid line shows the exponential decay $\propto
\exp(-cT)$ with $c = 2.0$~K$^{-1}$.}
\label{FanoTdepFitFig}
\end{figure}

In order to treat the thermal broadening quantitatively, we model a
simple quantum circuit for the present system as shown in
Fig.~\ref{FanoTdepFitFig} (a).  The QD is simply expressed as a
tunable resonator consisting of a tunnel barrier and a perfect
reflector.  The phase shift of the reflector is approximated to be
proportional to the gate voltage around the resonance.  The connecting
QW between the QD and the T-junction is simply a phase shifter of $kL$
($k$ : wave vector).  We put the S-matrix for the junction
$\boldsymbol{S}_{\rm T}$ as
\begin{equation}
\begin{pmatrix}
b_1\\ b_2\\ b_3
\end{pmatrix}
=\boldsymbol{S}_{\rm T}
\begin{pmatrix}
a_1\\ a_2\\ a_3
\end{pmatrix}
,
\;\;
\boldsymbol{S}_{\rm T}=
\begin{pmatrix}
\frac{1-a}{2} & -\frac{1+a}{2} & \sqrt{\frac{1-a^2}{2}}\\
-\frac{1+a}{2} & \frac{1-a}{2} & \sqrt{\frac{1-a^2}{2}}\\
\sqrt{\frac{1-a^2}{2}} & \sqrt{\frac{1-a^2}{2}} & a
\end{pmatrix}
,
\label{SmatrixT}
\end{equation}
to maintain the unitarity~\cite{ButtikerPRA1984}. $\{a_i\}$ and
$\{b_i\}$ are amplitudes of incoming and outgoing waves, as shown in
Fig.~\ref{FanoTdepFitFig} (a).  Here, we take $a$ as a real number,
which determines the direct reflection coefficient at the junction.
The S-matrix for the QW (phase shifter) is expressed as
\begin{equation}
\begin{pmatrix}
a_3\\ b_3'
\end{pmatrix}
=\boldsymbol{S}_{\rm QW}
\begin{pmatrix}
b_3\\ a_3'
\end{pmatrix}
,\;\;\;\;\;\boldsymbol{S}_{\rm QW}=
\begin{pmatrix}
0 & e^{i\beta}\\ e^{i\beta} & 0
\end{pmatrix},
\;\;\;\;\;\beta\equiv kL.
\end{equation}
The S-matrix for the tunnel barrier can then be written as
\begin{equation}
\begin{pmatrix}a_3'\\ a_4\end{pmatrix}
=\boldsymbol{S}_{\rm QD}
\begin{pmatrix}b_3' \\ b_4\end{pmatrix},
\;\;\;\;\;\boldsymbol{S}_{\rm QD}=
\begin{pmatrix}
\cos\phi & i\sin\phi\\ i\sin\phi & \cos\phi
\end{pmatrix}.
\end{equation}
Lastly, the reflector with a variable phase shift of $\theta$ is
simply expressed as
\begin{equation}
b_4=e^{i\theta}a_4.
\end{equation}

By calculating the combined S-matrix, the complex transmission
coefficient of the system is obtained as
\begin{equation}
t=\frac{1+a}{2}\cdot\frac{-1-e^{i(\theta+2\beta)}+(e^{2i\beta}+e^{i\theta})\cos\phi}
{1+ae^{i(\theta+2\beta)}-(ae^{2i\beta}+e^{i\theta})\cos\phi},
\label{ComplexTransCoef}
\end{equation}
where
\begin{equation}
\beta=kL=k_{\rm F}L+\frac{L}{\hbar v_{\rm F}}(\epsilon-\mu),
\end{equation}
since $k^2 = k_{\rm F}^2 + \frac{2m}{\hbar^2} (\epsilon-\mu)$ with
$\vert (k-k_{\rm F}) / k_{\rm F}\vert \ll 1$ ($m$ : effective mass of
electron).  Here, $\mu$ is the position of the chemical potential and
$\mu \equiv \alpha V_g$.  $L$, $k_{\rm F}$, and the Fermi velocity
$v_{\rm F}$ are estimated from the experimental conditions. The
resonance of the QD occurs at $\theta=0$, hence $\theta$ is taken as
$b(\epsilon-\epsilon_0)$, where $\epsilon_0 \equiv \alpha V_0$ as in
Eq.~\eqref{CondFanoEqn2}. It is easy to see that $|t|^2$ numerically
reproduces the first term of Eq.~\eqref{CondFanoEqn} near the
resonance.

The conductance of the system at the low bias ($\ll kT$) can be
expressed by the Landauer-B{\"u}ttiker formula as~\cite{KangPRB2001}
\begin{equation} 
G_{tot} = A \int d\epsilon |t|^2
\frac{1}{4kT}\cosh^{-2}(\frac{\epsilon - \mu}{2kT})+G_{bg}.
\label{CondFanoEqn3}
\end{equation}
We treat $a$, $b$, $\phi$, $V_0$, and $G_{bg}$ as fitting
parameters. While $V_0$ and $G_{bg}$ are slightly dependent on
temperature due to the neighboring Fano resonances, $a$, $b$ and
$\phi$ are treated as temperature-independent.  $A$ is left
temperature-dependent to absorb decoherence other than thermal
broadening.

We found that the fitting is insensitive to the value of $a$ as long
as it is smaller than $\sim 0.3$, which suggests that the specific
form of $\boldsymbol{S}_{\rm T}$ in Eq.~\eqref{SmatrixT} does not
affect the generality. Figure~\ref{FanoTdepFitFig} (b) shows the
results of the successful fitting for $a=0$.  Note that the number of
crucial fitting parameters is very small since most of the parameters
can be uniquely determined at the lowest temperature and taken as
temperature-independent.  Amplitude $A$ is shown in
Fig.~\ref{FanoTdepFitFig} (c) as a function of temperature. At 600~mK,
the amplitude $A$ still remains $\sim 40$~\% of that at the lowest
temperature. The observed strong temperature dependence, therefore, is
mostly due to the thermal broadening in the QW. If we set $L$ to zero,
the temperature dependence of Eq.~\eqref{CondFanoEqn3} would be much
weakened.

Interestingly, the behavior of $A$ is well fitted to $\propto
\exp(-cT)$ with $c = 2.0$~K$^{-1}$ between 30~mK and 500~mK.  Such
temperature dependence of the coherence is reminiscent of that in an
AB ring with the local and nonlocal
configurations~\cite{KobayashiJPSJ2002}, where the temperature
dependence of the AB amplitude was found to be weaker in the nonlocal
setup than in the local setup.  Theoretically it was pointed out that
the difference in the impedance of the probes seen from the sample is
important~\cite{SeeligPRB2003}.  In the present case, since one end of
the QD is cut and the nonlocal effect is observed, the impedance seen
from the sample (namely, the QD and the QW connecting it to the lead)
is very high and the situation is basically the nonlocal one they
treated.  Hence, the discussion in Ref.~[\onlinecite{SeeligPRB2003}]
might be applicable here.  In the present case, however, we do not
have the data that corresponds to the ``local'' setup, which can be
compared with the present ones.

\subsection{Phase Measurement of Electrons at a QD}
\label{subsec2}

Next, we discuss the application of the present geometry to the
measurement of the phase evolution at the QD. While the QD-AB-ring
geometry has been used for this
purpose~\cite{YacobyPRL1995,kats1996,SchusterNature1997,%
vanderWielScience2000,JiScience2000,JiPRL2002}, the T-coupled QD
should also provide information on the phase shift by the QD.  Since
in Eq.~\eqref{ComplexTransCoef}, the dip structure is mainly due to
the resonance and phase shift in the QD, we can, in principle, extract
information on the phase shift.  If we restrict ourselves around the
resonance point, we can utilize the simple Fano formula
[Eq.~\eqref{CondFanoEqn}], instead of the complicated analysis by
using the quantum circuit in the previous section.  To observe this,
we made the gate $V_L$ slightly open and allowed electrons to pass
through the QD. The system is now a QD-AB-ring system rather than a
T-coupled QD and the Fano effect in both the reflection mode and the
transmission mode is expected to occur.

\begin{figure}[tb]
\center \includegraphics[width=0.95\linewidth]{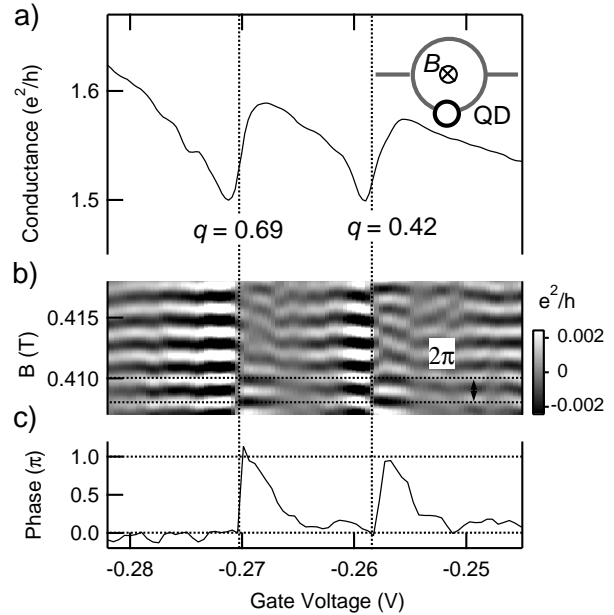}
\caption{(a) Two resonance dips in the conductance as a function of
$V_g$ at 30~mK at $B=0.405$~T.  The system now allows electrons to
pass through the QD.  The vertical lines indicate the positions of the
discrete energy levels in the QD. Note that the $q$ values are
positive.  (b) Gray-scale plots of the AB oscillation component of the
system as a function of $V_g$ and $B$. (c) Phase of the AB oscillation
obtained at each $V_g$.}
\label{TFanoAB}
\end{figure}

Figure~\ref{TFanoAB}~(a) shows the conductance of the system as a
function of $V_g$, where two resonance dips showing Fano line shape
are plotted. The dashed lines indicate the positions of the discrete
energy levels in the QD that are obtained by the aforementioned
fitting procedure. The values of the asymmetric parameter $q$ are
given in the figure. Note that the direction of the asymmetric tail,
namely, the sign of $q$, is the same for both dips.

Because the conduction through the lower arm is maintained very low,
these Fano features change minimally with the slight variation of $B$,
although there exists a coherent component in the transmission through
the QD, which appears as a small oscillation in conductance versus $B$.
This is the AB oscillation whose period is 2.0~mT as expected from the
ring size.  The AB oscillation was measured at each $V_g$ and the AB
component was extracted by fast Fourier transformation
(FFT). Figure~\ref{TFanoAB}~(b) shows the obtained AB component in the
gray-scale plot as a function of $V_g$ and $B$. The coherent component
through the QD is only of the order of $\sim 0.005 e^2/h$, in clear
contrast with the net signal up to the order of $\sim 0.08 e^2/h$
[Fig.~\ref{TFanoAB}~(a)].  This ensures that the Fano effect in the
reflection mode still plays a central role due to the slight opening of
the gate $V_L$. Contrastingly, in the previous
experiment~\cite{KobayashiPRL2002,KobayashiPRB2003} both $V_L$ and $V_R$
were made appropriately transmissible, and the amplitude of the AB
oscillation was in the same order of the net signal (see Fig. 8 in
Ref.~\onlinecite{KobayashiPRB2003}), which is a sign of the Fano effect
in the transmission mode.

Now, we focus on the behavior of the phase of the AB oscillation.  We
traced the conductance maximum as a function of $V_g$ and plotted in
Fig.~\ref{TFanoAB}~(c) in the unit of the AB period.  The phase
abruptly changes by $\pi$ just at the resonance points.  Such a phase
jump that occurs in the energy scale much smaller than $\varGamma$
reflects the two-terminal nature of the present QD-AB-ring system.
The AB oscillations at both dips are in-phase.  This is consistent
with the result of the previous report~\cite{YacobyPRL1995}.  Although
the origin of the in-phase nature of the Coulomb peak is still under
debate despite of several theoretical
studies~\cite{YeyatiPRB1995,BruderPRL1996,OregPRB1997,WuPRL1998,%
HackenbroichEPL1997,LeePRL1999,TaniguchiPRB1999}, it is clear that the
sign of $q$ is the same for both dips reflecting the in-phase nature.
This result, which is reproduced for other peaks in the present
experiment, demonstrates that ``phase measurement'' can be performed
simply by observing the sign of $q$, without investigating the AB
oscillation.  Interestingly, there also exists another slow $\pi$
phase shift away from the resonance point commonly for these peaks,
leading to their in-phase nature. Such behavior is consistent with
that observed in the Fano effect in the transmission
mode~\cite{KobayashiPRL2002,KobayashiPRB2003}.

\section{Conclusion}
We have realized the Fano effect in a QW with a side-coupled QD. The
temperature dependence of the resonance structure was analyzed by
including the thermal broadening, which turned out to be mostly
responsible for the rapid smearing of the resonances.  We have shown
that the phase measurement of electrons at a QD is also possible in this
geometry and gives a result consistent with that by a QD-AB-ring
system. While such a simple geometry as the T-coupled QD has never been
investigated experimentally, its clear advantage lies in that only a
coherent signal associated with the QD is obtained.  This work proves
that the Fano effect in this interferometer can be a powerful tool for
measuring the coherence and phase of electrons.

\section*{ACKNOWLEDGMENTS}

We thank F. B. Anders, M. Eto, T. Ihn, J. K{\"o}nig, and A. Schiller
for helpful comments.  This work is supported by a Grant-in-Aid for
Scientific Research and by a Grant-in-Aid for COE Research (``Quantum
Dot and Its Application'') from the Ministry of Education, Culture,
Sports, Science, and Technology of Japan. K.K. is supported by a
Grant-in-Aid for Young Scientists (B) (No.~14740186) from Japan
Society for the Promotion of Science.

\end{document}